\documentclass[conference]{IEEEtran}

\usepackage{cite}

\ifCLASSINFOpdf
   \usepackage[pdftex]{graphicx}
\else
\fi
\usepackage[cmex10]{amsmath}
\usepackage{algorithmic,algorithm}

\usepackage{array}

\usepackage[tight,footnotesize]{subfigure}

\newcommand{\MYfooter}{\smash{\scriptsize
\hfil\parbox[t][\height][t]{\textwidth}{\centering
This paper has been accepted for presentation in 4th  IEEE International Conference on Cloud Networking (IEEE CloudNet 2015) \\ 
to be held on 5-7 October, 2015, Niagara falls, Canada. This is an author copy. The respective copyrights are with IEEE.  }\hfil\hbox{}}}

\makeatletter

\def\ps@headings{%
\def\@oddhead{\MYfooter}
\def\@evenhead{\MYfooter}
}

\def\ps@IEEEtitlepagestyle{%
\def\@oddhead{\MYfooter}%
\def\@evenhead{\MYfooter}%
}

\makeatother

\begin{document}

\title{On Optimizing Replica Migration in Distributed Cloud Storage Systems}

\author{
    \IEEEauthorblockN{Amina Mseddi\IEEEauthorrefmark{1},  Mohammad Ali Salahuddin\IEEEauthorrefmark{1}\IEEEauthorrefmark{3}, Mohamed Faten Zhani\IEEEauthorrefmark{2}, Halima Elbiaze\IEEEauthorrefmark{1}, Roch H. Glitho\IEEEauthorrefmark{3} }
    \IEEEauthorblockA{\IEEEauthorrefmark{1}Universit\'e du Qu\'ebec \`A Montr\'eal, Montreal, Quebec, Canada \\}
    \IEEEauthorblockA{\IEEEauthorrefmark{2}\'Ecole de Technologie Sup\'erieure, Montreal, Quebec, Canada \\}
		\IEEEauthorblockA{\IEEEauthorrefmark{3}Concordia University, Montreal, Quebec, Canada \\}
    mseddi.amina@courrier.uqam.ca, mohammad.salahuddin@ieee.org, \\ mfzhani@etsmtl.ca, elbiaze.halima@uqam.ca, glitho@ciise.concordia.ca
}

\maketitle

\begin{abstract}
With the wide adoption of large-scale Internet services and big data, the cloud has become the ideal environment to satisfy the ever-growing storage demand, thanks to its seemingly limitless capacity, high availability and faster access time. In this context, data replication has been touted as the ultimate solution to improve data availability and reduce access time. However, replica placement systems usually need to migrate and create a large number of data replicas over time between and within data centers, incurring a large overhead in terms of network load and availability. 
In this paper, we propose CRANE, an effiCient Replica migrAtion scheme for distributed cloud Storage systEms. CRANE complements any replica placement algorithm by efficiently managing replica creation in geo-distributed infrastructures by (1) minimizing the time needed to copy the data to the new replica location, (2) avoiding  network congestion, and (3)~ensuring a minimal availability of the data. Our results show that, compared to swift (the OpenStack project for managing data storage), CRANE is able to minimize up to 30\% of the replica creation time and 25\% of inter-data center network traffic, while ensuring the minimum required availability of the data. 

\end{abstract}

\IEEEpeerreviewmaketitle

\section{Introduction}
With the wide adoption of large-scale Internet services and big data, the cloud has become the ultimate resort to cater to the ever-growing demand for storage, providing seemingly limitless capacity, high availability and faster access time.
Typically, cloud providers build several large-scale data centers in geographically distributed locations. Then, they rely on data replication as an effective technique to provide fault-tolerance, reduce end-user latency and minimize the amount of data exchanged through the network. As a result, replica management has become one of the major challenges for~cloud~providers.

In recent years, a large body of work has been devoted to several challenges related replica management in distributed cloud storage systems. A large part of the research efforts were mostly dedicated to replica placement problem, considering different goals such as minimizing storage costs, improving fault-tolerance and access delays \cite{gfs, hdfs,ReplicaDynamic, ReplicaDynamicCDRM, ReplicaPlacement}.
However, replica placement systems may result in a huge number of data replicas created or migrated over time between and within data centers, incurring large amounts of traffic between data centers. This can be the case in different scenarios: for instance, when a new data center is added to the cloud provider's infrastructure, when a data center is scaled up or down, when recovering from a disaster or simply when achieving performance or availability goals, requiring the creation and the relocation of~a~large~number~of~replicas. 

Naturally, several impacts may be expected when such large data bulk transfer of replicas is triggered. These impacts can be summarized as follows:

$\bullet$ As copying data consumes resources (e.g., CPU, memory, disk I/O) at both the source and the destination machines, these nodes will experience more contention for the available capacity, which may slow down other tasks running on them.
	
$\bullet$ Recent research revealed that traffic exchanged between data centers account for up to $45\%$ of the total traffic in the backbone network connecting them \cite{InterdatacenterTraffic}. This ever-growing exchange of tremendous amounts of data between data centers may overload the network, especially when using the same paths or links. This can hurt the overall network performance in terms of latency and also packet loss. 
	
	
$\bullet$ Replica migration processes are usually distributed and asynchronous as is the case for Swift, the OpenStack project for managing data storage \cite{Swift1}. That~is,~when a replica is to be relocated or created in a new destination machine, every machine in the infrastructure already storing the same replica will try to copy the data to the new destination. There is no coordination or synchronization between the sending nodes. This will not only lead to unneeded redundancy as the same data is copied from different sources at the same time, but will also further exacerbate the congestion in the data center network.
		
	
$\bullet$ Replicas are usually placed in geographically distributed locations, so as to increase data availability over time and reduce user-perceived latency. When a replica have to be created/migrated in a new location, it  will not be available until all its content is copied from other existing replicas. If this process takes too long, it might hurt the overall data availability, if the number of available replicas is not sufficient to accommodate all user requests. For instance, in order to ensure availability, Swift ensures that at least two replicas of the data are available at any point in time (according to the default configuration \cite{Swift1}).

In order to alleviate all the aforementioned problems, it~is critical to make sure that replicas are created as soon as possible in their new locations without inferring network congestion or high creation time. 
This requires to carefully select the source replica from which the data will be copied, the paths through which the data will be sent and the order in which replicas are created.

To address these challenges, we start by formulating the replica migration/creation problem as an Integer Linear Program (ILP). We then propose (CRANE\footnote{CRANE:(Mechanical Engineering) a device for lifting and moving heavy objects, typically consisting of a moving boom, beam, or gantry from which lifting gear is suspended }) an effiCient Replica migrAtion scheme for distributed cloud Storage systEms. 
CRANE is a novel scheme that manages replica creation in geo-distributed infrastructures with the goal of (1) minimizing the time needed to copy the data to the new replica location, (2) avoiding  network congestion, and (3) ensuring a minimal availability for each replica. CRANE can be used with any existing replica placement algorithm in order to optimize the time to create and copy replicas and to minimize resources needed to reach the new placement of the replicas. In addition, it ensures that at any point in time, data availability is above a predefined minimum value.

The rest of the paper is organized as follows. Section \ref{sec:RelatedWork} surveys the related work on replica placement and migration in the cloud. Section \ref{sec:MotivatingExample} presents an example illustrating how different replica creation and migration strategy can impact performance metrics. We then formulate the replica creation problem in Section \ref{sec:TheReplicaMigrationProblem}. Our proposed solution is presented in Section \ref{sec:SolutionDesign}. Finally, we provide  in Section \ref{sec:PerformanceEvaluation} some simulation results that assess the performance of CRANE and compare it to Swift and~we~conclude in Section \ref{sec:Conclusion}. 

\section{Related Work}
\label{sec:RelatedWork}

In this section, we survey relevant works on replica management in the cloud. 
Several efforts have been devoted to put forward effective solutions for replica placement problem. That is to determine the optimal number and placement of data replicas in order to 
achieve several objectives such that minimizing hosting costs, reducing access delay to the data, and maximizing data availability \cite{gfs, hdfs, ReplicaDynamic,ReplicaDynamicCDRM, ReplicaPlacement, Zaman2011}. Once the replica placement algorithm is executed, a new placement of replicas is determined in order to achieve the desired objective. Hence, some existing replicas should be torn down and some new ones should be created across the infrastructure. 

In this work, we do not focus on the replica placement but rather on reducing the overhead of migrating from an original placement of replicas to the new one, which should take place right after the execution of the replica placement algorithm. In the last few years, very few proposals have looked at this problem but overlooked many important parameters. For instance, Kari~et~al.~\cite{kari2011data} proposed a scheme that tries to find an optimal migration schedule for data in order to minimize the total migration time. They take into account the heterogeneity of storage nodes in terms of the number of simultaneous transfers they can handle. However, they have overlooked availability requirements as well as network-related constraints such as bandwidth limits and propagation delays. Other works \cite{MigrationCloning, MigrationCloning2} proposed different approximation algorithms to solve the problem; however they always aim at minimizing migration times without considering the availability of data during the migration process. 
Finally, Swift, the OpenStack project for managing data storage \cite{Swift1}, implements a data replica placement algorithm along a replica migration one. As a placement strategy, blocs of data (called hereafter as partitions) are replicated and distributed across the distributed infrastructure according to the as-unique-as-possible algorithm \cite{DataPlacementSwift}, which ensure that partition's replicas are physically stored as far as possible from each other in order to ensure high availability. In terms of replica creation and migration, Swift simply do migrates replicas between data centers without considering the network available capacity. However, it ensures a high availability of the data by allowing only one migration per partition each time interval (usually, one hour) so that only one replica can be missing at a particular point of time. Of course, the1-hour waiting time for triggering migrations will significantly increase the total time needed to reach the new optimal placement of replicas.


Different from previous work, CRANE takes into account not only the network available capacity but also data availability during the creation of the new replicas. Furthermore, it capitalizes on the existence of multiple replica across the network in order to carefully select the source of the data and the transmission path in order to avoid network congestion and minimize data migration time.

\section{Motivating Example}
\label{sec:MotivatingExample}

\begin{figure}[h!]
\begin{center}
\subfigure[Initial replication mapping]{
\includegraphics[scale=0.2]{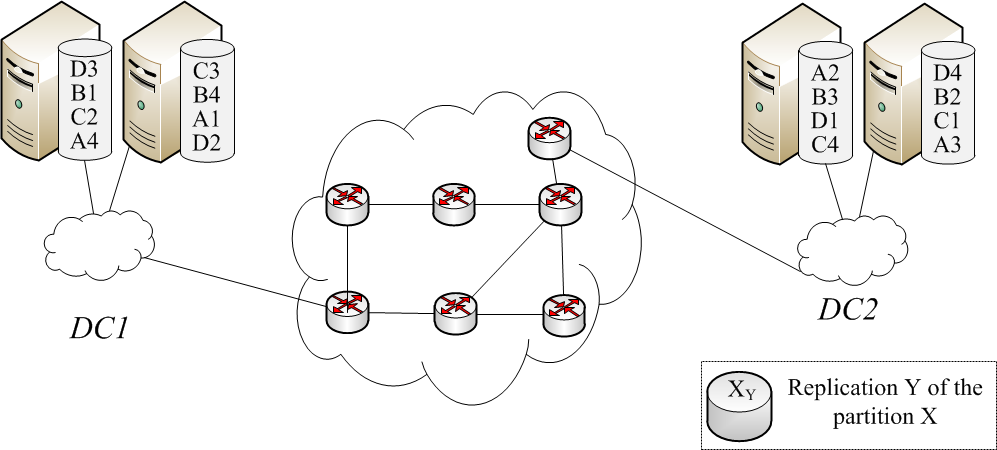}
\label{InitialMapping}}
\hfill
 \subfigure[Final replication mapping]{
 \includegraphics[scale=0.2]{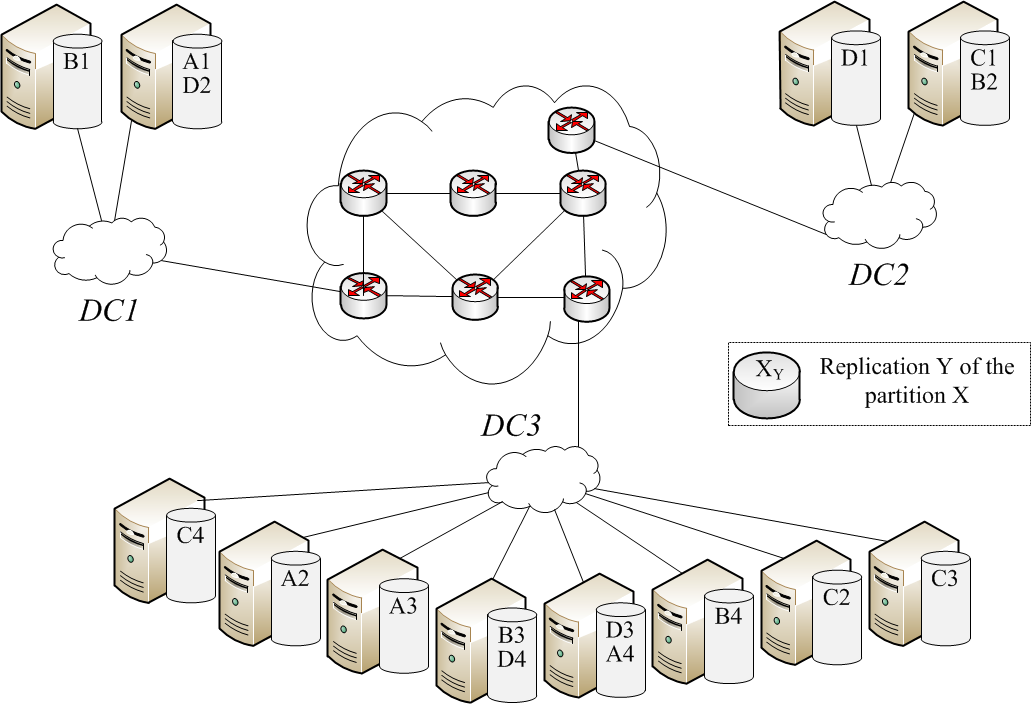}
\label{FinalMapping}}
\caption{Initial and final replication mapping}
\label{Mapping}
\end{center}
\end{figure}

To introduce our proposed replica placement solution, a motivating example is described in this section. 
Let's consider a cloud system composed of two data centers that span multiple geographic regions. The data centers have different storage capacities and are interconnected by different capacity links. This cloud deployment uses Swift as a distributed solution for managing storage functionalities. Consider a scenario where 4 partitions A, B, C and D with sizes 300 GB, 100 GB, 500 GB and 200 GB, respectively, are configured. Each partition is replicated 4 times and the replicas are stored across data centers. Fig. \ref{InitialMapping} illustrates the initial mapping of the replicas on the data centers.

When a new data center is added, partitions are relocated according to the as-unique-as-possible algorithm, respecting the capacity of the disks on servers in the data centers. To reach the new configuration, bulk volumes of replications will be migrating. Starting from the time where we decide to relocate the replications, all the Swift components will operate according to the new mapping of the replications, directing client requests to non-existent replicas, resulting in unavailability. Swift requires a majority of replicas responding to consider a retrieval operation successful. Thus, to avoid unavailability of the replicas, Swift specifies in its configuration the number of hours to restrict moving a partition more than once. In our scenarios, we assume this parameter to be one hour. Thus, Swift will not move two replicas of the same partition at the same time, but wait an hour before moving the second replica. It is important to note that the second migration is not triggered automatically. Instead, the cloud administrator should run a command to recompute the location of the replications, and then migration starts. Also, in this scenario, the minimum tolerated availability of replicas of each partition is assumed to be $\dfrac{3}{4}$. Fig. \ref{FinalMapping} shows the optimal locations of the different replications according to the disk capacities and the as-unique-as-possible algorithm. The depicted final configuration is reached after several calculations of locations and migrations.
\begin{figure}[h!]
\begin{center}
\includegraphics[scale=0.3]{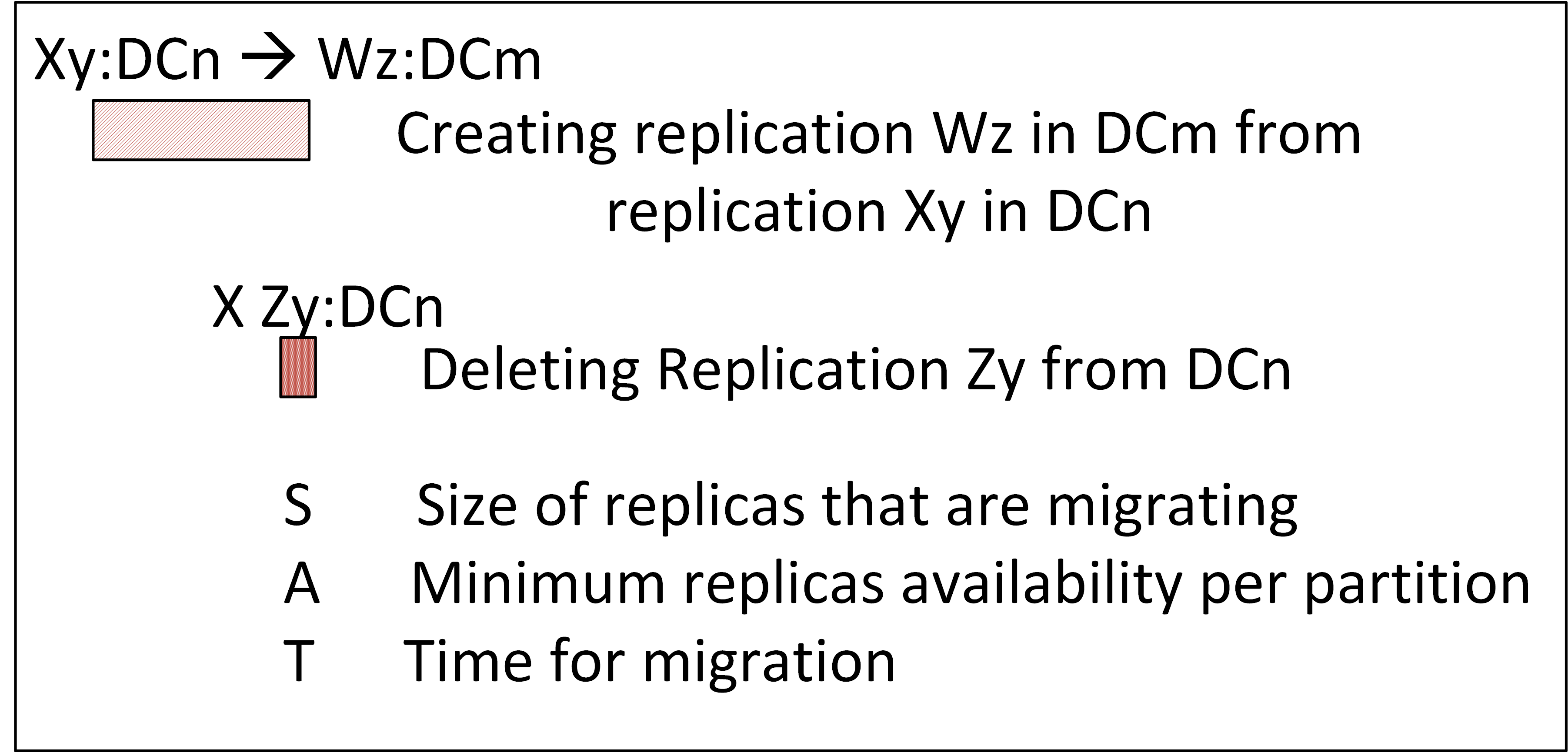}
\hfill
\subfigure[Example of Swift replication migration sequence]{
\includegraphics[scale=0.22]{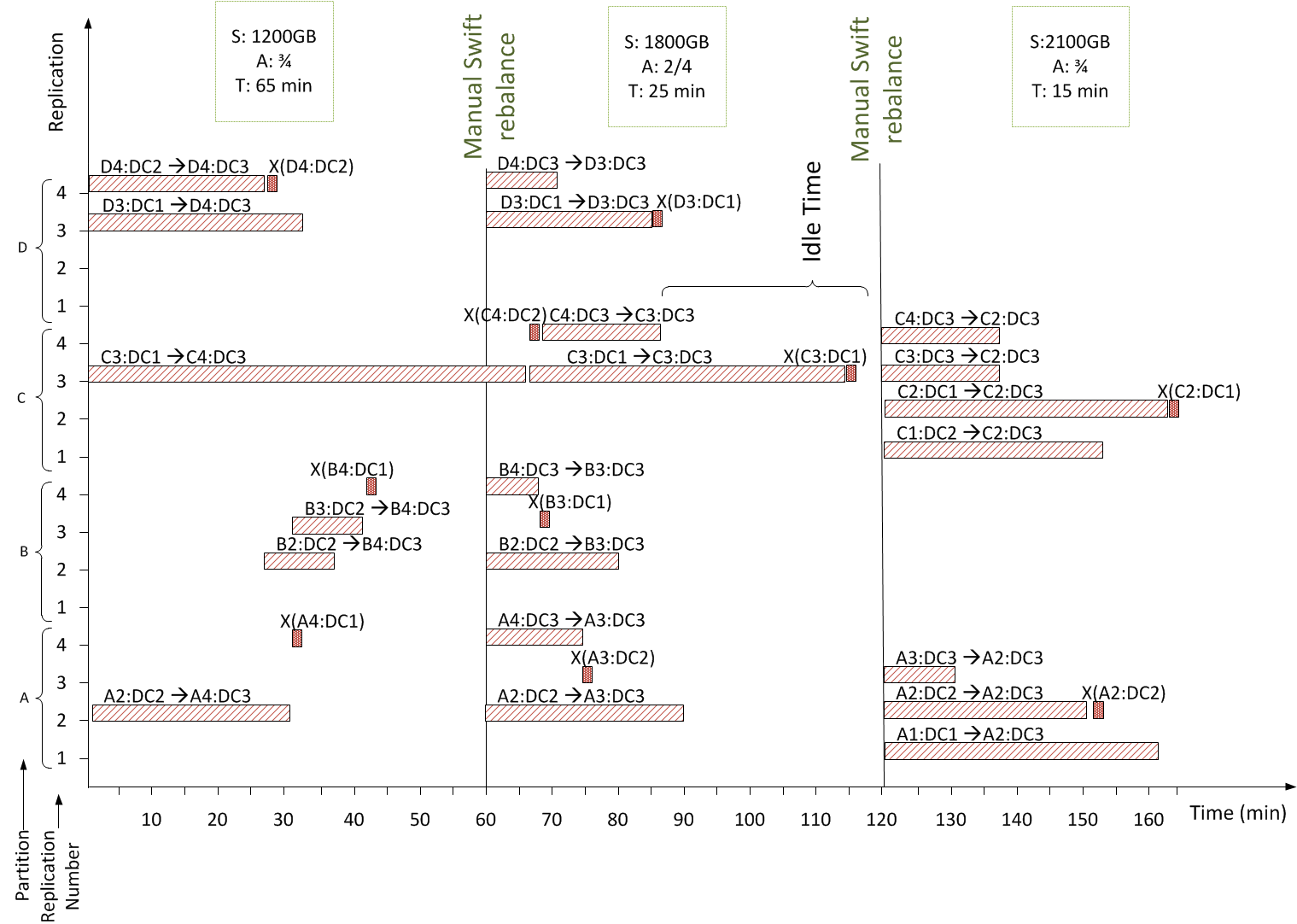}
\label{Sequence}}
\hfill
 \subfigure[Novel replication migration sequence]{
 \includegraphics[scale=0.21]{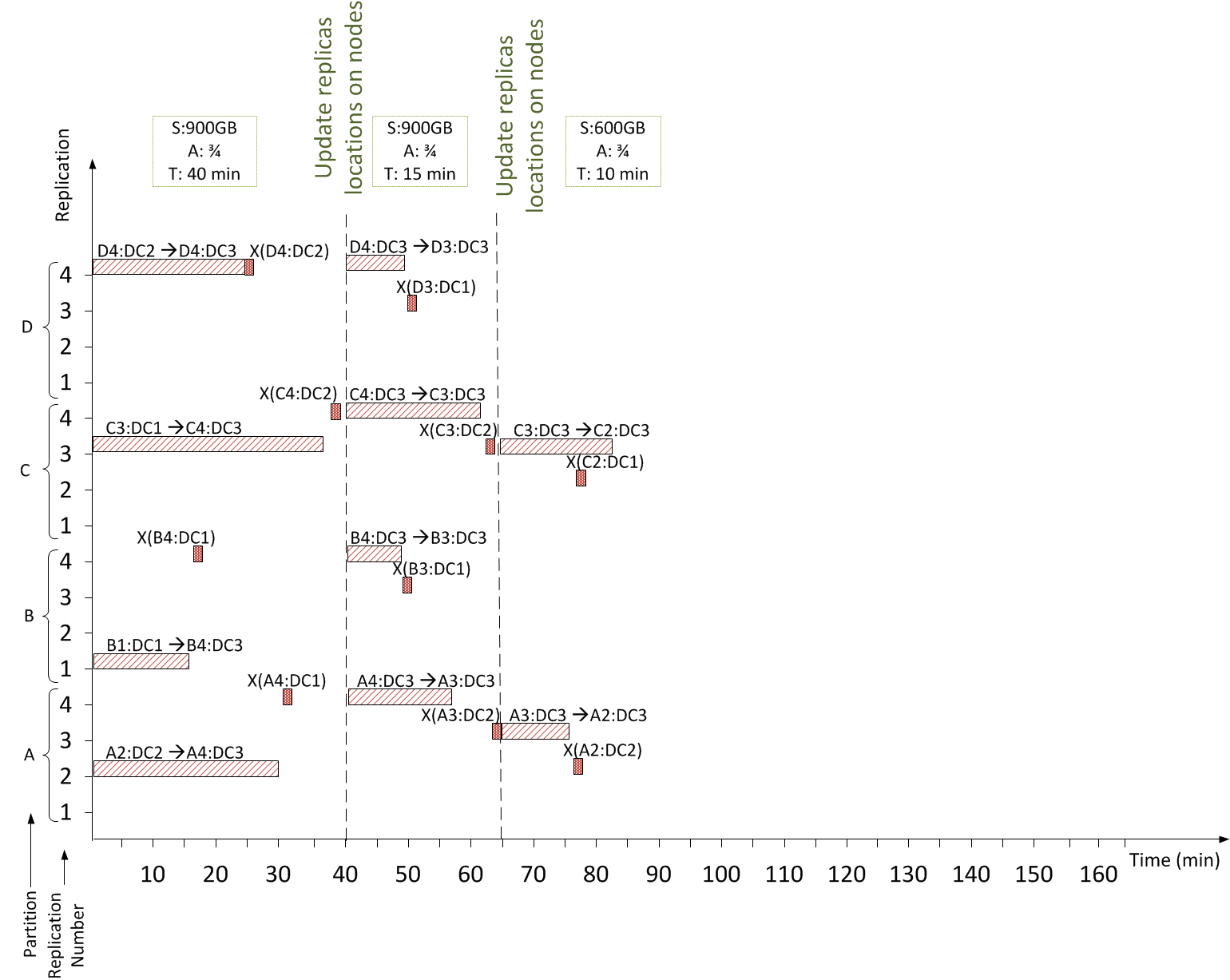}
\label{SequenceOptimal}}
\caption{Replication migration sequences}
\label{Sequences}
\end{center}
\end{figure}

Migration of replications in Swift is a peer-to-peer asynchronous process. Each storage server will sequentially run through its replications and compare them to the ones on the servers assigned to hold a copy of this replication. The migration process uses a push model, so each time a replication needs to be present in one server and it's not, this replication will  be copied from the local server to the remote one. And if a server holds a copy of one replication that shouldn't be there it should delete it. Fig. \ref{Sequence} represents an example of the replication migration sequence. In this figure, the bottom line represents time in minutes and the replication identifiers of each partition are represented on the vertical line. For instance each storage server starts sequentially copying the replicas to the servers that should hold a copy of them and are not. However, the following problems arise:

\emph{  Availability:} The number of hours to restrict moving a replica of the same partition is a parameter defined by the storage provider. In this scenario it's set to one hour. But migration of replications might take more than one hour. This may have a bad impact on the minimum tolerated availability of replicas per partition. In the depicted scenario, copying replica C3 from DC1 to create C4 on DC3 have taken more than one hour, and the new computation of location of replicas dictated that D3 should be migrated to DC3. So, in the second migration sequence, we have both C4 and C3 that are not available on DC3 during the five first minutes. This violates the minimum required replicas available per partition.

\emph{Redundancy:} There is a redundancy on the creation of the replica D4 on DC3. Infact, replication D3 and D4 are copied from DC1 and DC2, respectively. This redundancy implies needless bandwidth consumption, which increases replication migration time. Moreover, the copying of B2 from DC2 to DC3 is delayed. Indeed, all the storage servers were copying the first replica in their list, that is, B2 started to be copied only when the server was finished copying D4.

\emph{Migration time:} Each storage server on Swift starts copying the replicas of the partitions that need to be migrated without any calculation on the delays that this task would take. Therefore, one storage server could end up copying a replica, even though, there exists another server with a replica of the same partition, which could have performed the replication faster. For example, in order to create D4 in DC3, copying D3 from DC1 takes less time than copying D4 from DC2. This could be due to the difference on the available bandwidth on the links and the propagation delays between data centers. Moreover copying a replica within the same data center will lead to better migration time.

 \emph{Idle time:}  We notice there is always an idle time between two sequences of migration.

To avoid the aforementioned drawbacks, we propose an enhanced migration solution illustrated in Fig. \ref{SequenceOptimal}. Note, avoiding the redundancy of copying replicas increases available bandwidth and thus decreases migration time. Also, smart selection of copying source further reduces the migration time significantly. In fact, creating replicas from within the same data center or from a distant data center with links having better propagation delay and less bandwidth consumption will lead to a faster migration time. Furthermore, the minimum replica availability tolerated per partition should be maintained. Finally, automating the recalculation of the location after each sequence of migration leads to avoiding idle time and faster convergence to the final mapping of replicas in data centers.

\section{The Replica Migration Problem}
\label{sec:TheReplicaMigrationProblem}

\subsection{Problem Statement}
Given a network represented by a graph$\ G=(S,\ E)$, where $S=\left\{s_1, s_2, ..., s_i, ..., s_k, ..., s_{\left\vert{}S\right\vert{}}\right\}$ is the set of all servers across data centers. We assume that data centers are connected through a backbone network. The backbone's links are represented by a set of edges   $E$. Each edge $e\in{}E$ is characterized by a bandwidth capacity $\ B_e$. Let $P =\{p_{1}, p_{2}, ..., p_{j}, ..., p_{|P|} \}  $ denote the set of partitions, replicas of which are stored across the servers where  $|p_{j}|$ is the size of replica of partition $p_{j}$. We define a configuration as a particular placement of the replicas of partitions within servers. Given an initial and a final configuration, denoted respectively by $C^I$ and $C^F$, our goal is to find the optimal sequence of replica migrations that minimizes the migration time from $C^I$ to  $C^F$ while meeting the minimum partition availability threshold $A$ and abiding by edge bandwidth capacities. We model this as an Integer Linear Programming (ILP) problem.
Tables I and~II~show respectively the inputs of the ILP and its variables.
\begin{table} [!htb]
\caption{Problem inputs}
\begin{tabular}{|m{23pt}|m{208pt}|} 
\hline
	\textbf{Input}
 	& 
	\textbf{Definition}
 \\
\hline
	$S$
 	& 
 	Set of servers across data centers, where $S=\left\{s_1,s_2, ...,s_i, ..., s_k, ..., s_{\left\vert{}S\right\vert{}}\right\}$
\\ 
 
\hline
	$E$
 	& 
 	Set of edges connecting servers in $S$
\\

\hline
	$B_e$
	& 
	Bandwidth capacity $\forall{}e\in{}E$
\\

\hline
 	$P$
	& 
	Set of partitions, where $P=\left\{p_1,p_2, ...,p_j, ...,p_{\left\vert{}P\right\vert{}}\right\}$ and $\left\vert{}p_j\right\vert{}$ is the size of partition $p_j$ 
 \\

\hline
	$C^I$
	& 
	$
	\left\vert{}S\right\vert{}\times{}\left\vert{}P\right\vert{}
	$ matrix representing an initial configuration, where
	$c_{i,j}^I=\left\{\begin{array}{l}1,\ \mbox{if replica of}\ p_j\ \mbox{is}\ \mbox{stored}\ \mbox{on}\ s_i\ \  \\
	0,\ \mbox{otherwise}\end{array}\right.$

\\

\hline
	$C^F$
	&
	$
	\left\vert{}S\right\vert{}\times{}\left\vert{}P\right\vert{}
	$ matrix representing a final configuration, where $c_{i,j}^F=\left\{\begin{array}{l}1,\ \mbox{if replica of}\ p_j\ \mbox{is}\ \mbox{stored}\ \mbox{on}\ s_i\ \  \\
0,\ \mbox{otherwise}\end{array}\right.$
\\

\hline
	$T$
 	&  
	Worst-case migration time
 \\

\hline
	$G$
	&
	$
	\left\vert{}S\right\vert{}\times{}\left\vert{}S\right\vert{}\times{}\left\vert{}E\right\vert{}$ matrix representing edges used in a path,  where \\ & 
$g_{i,k,e}=\left\{\begin{array}{l}1,\ \mbox{if}\ \mbox{edge}\ e\ \mbox{is}\ \mbox{used}\ \mbox{in}\ \mbox{shortest}\\ \ \  \ 
\mbox{path}\  \mbox{between}\ s_i\ \mbox{and}\ s_k\  \\
0,\ \mbox{otherwise} \end{array}\right.$
 \\

\hline
$\beta{}$
&  
A big constant 
 \\
\hline
\end{tabular}
 \end{table}

\begin{table}[!htb]

\caption{Problem variables}
\begin{tabular}{|p{22pt}|p{208pt}|}
\hline
\parbox{22pt}{\raggedright 
\textbf{Variable}
} & \parbox{208pt}{\raggedright 
\textbf{Definition}
} \\
\hline
\parbox{22pt}{\raggedright 

$X$

} & \parbox{208pt}{\raggedright 

$\left\vert{}S\right\vert{}\times{}\left\vert{}P\right\vert{}\times{}\left\vert{}S\right\vert{}\times{}T$ matrix representing migration sequence, where
$x_{i,j,k,t}=\left\{\begin{array}{l}1,\ \mbox{if}\ s_i\ \mbox{is migrating replica of}\ p_j\ \mbox{to}\ s_k\
\mbox{at time}\ t\  \\
0,\ \mbox{otherwise}\end{array}\right.$
} \\
\hline
\parbox{22pt}{\raggedright 

$Y$

} & \parbox{208pt}{\raggedright 
$
\left\vert{}S\right\vert{}\times{}\left\vert{}P\right\vert{}\times{}\left\vert{}S\right\vert{}$ matrix representing a need for partition migration, where
$y_{i,j,k}=\left\{\begin{array}{l}1,\ \mbox{if}\ s_i\ \mbox{is the provider of replica of}\ p_j\ \mbox{to}\ s_k
\\
0,\ \mbox{otherwise}\end{array}\right.$
} \\
\hline
\parbox{22pt}{\raggedright 

$Z$

} & \parbox{208pt}{\raggedright 

$\left\vert{}S\right\vert{}\times{}\left\vert{}P\right\vert{}\times{}T$ matrix representing replica placement, where
$z_{i,j,t}=\left\{\begin{array}{l}1,\ \mbox{if}\ s_i\ \mbox{has}\ \mbox{replica} \ \mbox{of} \ \ p_{j}\ \mbox{at time}\ t\  \\
0,\ \mbox{otherwise}\end{array}\right.$
} \\
\hline
\parbox{22pt}{\raggedright 
$L$

} & \parbox{208pt}{\raggedright 

$\left\vert{}E\right\vert{}\times{}T$ matrix representing load $l_{e,t}$ on edge $e$ at time $t$,  where $l_{e,t}<B_e$
} \\
\hline
\parbox{22pt}{\raggedright 

$R$

} & \parbox{208pt}{\raggedright 

$\left\vert{}S\right\vert{}\times{}\left\vert{}P\right\vert{}\times{}\left\vert{}S\right\vert{}\times{}T$ matrix, where $r_{i,j,k,t}$ represents the bandwidth allocated for migrating replica of $p_j$ from $s_i$ ro $s_k$ at time $t$
} \\
\hline
\parbox{22pt}{\raggedright 

$D$

} & \parbox{208pt}{\raggedright

$\left\vert{}S\right\vert{}\times{}\left\vert{}P\right\vert{}$ matrix representing the replicas that are to be deleted at time $T$, where 
$d_{i,j}=\left\{\begin{array}{l}1,\ \mbox{if replica of}\ \ p_j\ \mbox{will be deleted from} s_i\ 
\\
0,\ \mbox{otherwise}\end{array}\right.$
} \\
\hline
\parbox{22pt}{\raggedright 

$V$

} & \parbox{208pt}{\raggedright 

$\left\vert{}S\right\vert{}\times{}\left\vert{}S\right\vert{}\times{}T$
matrix, where $v_{i,k,t}\ $represents the capacity of the path between $s_i$ to
$s_k$ at time $t$
} \\
\hline
\parbox{22pt}{\raggedright 

$W$

} & \parbox{208pt}{\raggedright 
A vector of size $T$, where $w_t=\left\{\begin{array}{l}1,\ \mbox{if migration is
in progress at time}\ t\  \\
0,\ \mbox{otherwise}\end{array}\right.$
} \\
\hline
\end{tabular}
\end{table}
\subsection{Constraints}
Given the initial and final configurations, any discrepancy in the configurations necessitates either the migration or the deletion of the partition replicas. Consider that the migration or deletion of the replica is identified by variables
${\ y}_{i,j,k}$  and $d_{k,j}$, respectively. Then, if a server $\
s_k$ has replica of partition $p_j$ in the initial and final configuration and no action is required, then there should be no migration of replica of partition  $p_j$  from any server $s_{i}$ to $s_{k}$, that is, $\sum_{i=1}^{\vert{}S\vert{}}y_{i,j,k}=0$,  and there should be no deletion of replica of partition $p_{j}$ from server $s_{k}$, that is,  $d_{k,j}=0$, as in (\ref{eq:1}).  However, if the replica of partition $p_{j}$ is present on server $s_{k}$ in the initial configuration and not in the final configuration, then the partition should be eventually deleted, hence, $d_{k,j}$ is set to~1, with constraint (\ref{eq:2}). Most importantly, in constraint (\ref{eq:3}) we capture the need for a replica migration, if the server $s_{k}$ does not have the replica of partition $p_{j}$ in the initial configuration. In this case, there should be some server $s_{i}$ that delivers the replica of partition $p_{j}$ to server $s_{k}$, therefore $\sum_{i=1}^{\vert{}S\vert{}}y_{i,j,k}=1$.
\begin{equation}
c_{k,j}^I+\sum_{i=1}^{\left\vert{}S\right\vert{}}y_{i,j,k}-\ d_{k,j}=c_{k,j}^F \ \ \ \forall{}1\leq{}k\leq{}\left\vert{}S\right\vert{},1\leq{}j\leq{}\left\vert{}P\right\vert{}
\label{eq:1}
\end{equation}
\begin{equation}
d_{k,j}\geq{}\ c_{k,j}^I-\ c_{k,j}^F \ \ \ \forall{}1\leq{}k\leq{}\left\vert{}S\right\vert{},1\leq{}j\leq{}\left\vert{}P\right\vert{}
\label{eq:2}
\end{equation}
\begin{equation}
\sum_{i=1}^{\left\vert{}S\right\vert{}}y_{i,j,k}\geq{}c_{k,j}^F-c_{k,j}^I\ \ \forall{}1\leq{}k\leq{}\left\vert{}S\right\vert{},1\leq{}j\leq{}\left\vert{}P\right\vert{}
\label{eq:3}
\end{equation}
Before we can initiate the migration, we have to identify the servers $s_{i}$ that hold the replica of partition $p_{j}$ at time $t$, in variable $z_{i,j,t}$ in constraint (\ref{eq:4}). To begin the migration, only those servers $s_{i}$ can participate in the replica migration that hold a replica of partition $p_{j}$ in the initial configuration.
\begin{equation}
c_{i,j}^I\leq\beta{}\cdot{}z_{i,j,t} \ \ \ \ \ \forall{}1\leq{}i\leq{}\left\vert{}S\right\vert{},1\leq{}j\leq{}\left\vert{}P\right\vert{},1\leq{}t\leq{}T
\label{eq:4}
\end{equation}
The variable $z_{k,j,t}$ that indicates potential sources of replicas that indicates potential sources of replicas is updated in each time instance $t$, as servers $s_{k}$ may begin to hold a copy of the replica of partition $p_{j}$, due to migration in earlier time instances, as in constraints (\ref{eq:5}) and (\ref{eq:6}). A server holds a copy of the replica when the sum of all the bandwidth allocated, $r_{i,j,k,t'}$ to the migration of that replica of partition $p_{j}$ from source $s_{j}$ to destination $s_{k}$, in previous time instances $\forall t' < t$, equals the size of the partition $p_{j}$. Then, in following time instances, server $s_{k}$ could potentially participate in the replica migration.
\vspace{-5mm}
\begin{multline}
\left\vert{}p_j\right\vert{}-\sum_{k=1,i\not=k}^{\left\vert{}S\right\vert{}}\sum_{t^{'}=1}^tr_{i,j,k,t^{'}}\leq{}\beta{}\cdot{}\left(1-z_{k,j,t+1}\right)\ \\
\forall{}1\leq{}i\leq{}\left\vert{}S\right\vert{},1\leq{}j\leq{}\left\vert{}P\right\vert{},1\leq{}t\leq{}T-1
\label{eq:5}
\end{multline}
\vspace{-10mm}
\begin{multline}
\left\vert{}p_j\right\vert{}-\sum_{k=1,i\not=k}^{\left\vert{}S\right\vert{}}\sum_{t^{'}=1}^tr_{i,j,k,t^{'}}\geq{}1-z_{k,j,t+1} \\
\forall{}1\leq{}i\leq{}\left\vert{}S\right\vert{},1\leq{}j\leq{}\left\vert{}P\right\vert{},1\leq{}t\leq{}T-1
\label{eq:6}
\end{multline}
The bandwidth allocated for migration of replica of partition $p_{j}$ at time $t$ from source $s_{i}$ to destination server $s_{k}$, in variable $r_{i,j,k,t}$, is dependent on the capacity of the shortest path from $s_{i}$ to $s_{k}$, in variable $v_{i,k,t}$, or the remaining size of the partition replica to be migrated in constraint (\ref{eq:7}).
\vspace{-3mm}
\begin{multline}
r_{i,j,k,t}=argmin\{\left\vert{}p_{j}\right\vert{}-\sum_{k=1,i\not=k}^{\left\vert{}S\right\vert{}}\sum_{t'=1}^{t-1}r_{i,j,k,t'},v_{i,k,t}\}\ \\ \forall{}1\leq{}i,k,\
i\not=k\leq{}\left\vert{}S\right\vert{},1\leq{}j\leq{}\left\vert{}P\right\vert{},1\leq{}t\leq{}T
\label{eq:7}
\end{multline}
The capacity of the shortest path between $s_{i}$ and $s_{k}$ is inferred from the load on the edges traversed in the path. The edge with the least capacity, bounds the capacity of the path from above. The available path capacity at time $t$ is in constraint (\ref{eq:8}). The edges can be traversed by multiple paths, that is, multiple paths between different source destination pairs can share common edges. Therefore, the load on an edge, not exceeding edge capacity, is conjured as the sum of all partition replicas migrating in the network across all source and destination servers at time $t$, on edge $e$ in the shortest path between $s_{i}$ and $s_{k}$, by constraints (\ref{eq:9}).
\vspace{-3mm}
\begin{multline}
v_{i,k,t}=argmin\{\left(B_e-l_{e,t}\right)\cdot{}g_{i,k,e}\ 1\leq{}e\leq{}\left\vert{}E\right\vert{} \} \\  \forall{}1\leq{}i,k,\
i\not=k\leq{}\left\vert{}S\right\vert{},1\leq{}t\leq{}T
\label{eq:8}
\end{multline}
\vspace{-10mm}
\begin{multline}
l_{e,t}=\sum_{i=1}^{\left\vert{}S\right\vert{}}\sum_{j=1}^{\left\vert{}P\right\vert{}}\sum_{k=1}^{\left\vert{}S\right\vert{}}g_{i,k,e}\cdot{}r_{i,j,k,t}\ \ \forall{}1\leq{}e\leq{}\left\vert{}E\right\vert{},1\leq{}t\leq{}T
\label{eq:9}
\end{multline}
Once the model has been initialized with potential providers, need for migration and the bandwidth allocated for the migration of partition replicas, we can initiate migration by associating the replica of partition migration indicator variable $x_{i,j,k,t}$ with need for migration of replica of partition $p_{j}$ from $s_{i}$ to $s_{k}$, in variable $y_{i,j,k}$, in constraints (\ref{eq:10}) and (\ref{eq:11}). Consequentially, from (\ref{eq:10}) and (\ref{eq:11}), we ensure only sequential migration of replica of partition, since concurrency is set to~1. Furthermore, in constraint (\ref{eq:12}), we bind the source server $s_{i}$, such that, only those servers that hold complete replica of partition $p_{j}$ can initiate migration.  
\begin{equation}
\sum_{t=1}^Tx_{i,j,k,t}\leq{}\beta{}\cdot{}y_{i,j,k}  \ \ \ \ \forall{}1\leq{}i,k,\
i\not=k\leq{}\left\vert{}S\right\vert{},1\leq{}j\leq{}\left\vert{}P\right\vert{}
\label{eq:10}
\end{equation}
\begin{equation}
\sum_{t=1}^Tx_{i,j,k,t}\geq{}y_{i,j,k}   \ \ \ \  \forall{}1\leq{}i,k,\
i\not=k\leq{}\left\vert{}S\right\vert{},1\leq{}j\leq{}\left\vert{}P\right\vert{}
\label{eq:11}
\end{equation}
\begin{equation}
\sum_{k=1}^{\left\vert{}S\right\vert{}}x_{i,j,k,t}\leq{}z_{i,j,t}   \ \ \ \  \forall{}1\leq{}i\leq{}\left\vert{}S\right\vert{},1\leq{}j\leq{}\left\vert{}P\right\vert{},1\leq{}t\leq{}T
\label{eq:12}
\end{equation}
To ensure continuous sequential migration of replica of partition $p_{j}$, from same source server $s_{i}$ to same destination server $s_{k}$ for next time instance $t+1$, we set the indicator of migration is progress, in variable $x_{i,j,k,t+1}$, to 1, until the entire replica of the partition has been migrated. This is depicted in constraints (\ref{eq:13}) and (\ref{eq:14}). 
\vspace{-5mm}
\begin{multline}
{\beta{}\cdot{}x}_{i,j,k,t+1}\geq{}\left\vert{}p_j\right\vert{}-\sum_{t^{'}=1}^tr_{i,j,k,t^{'}} \\ \forall{}1\leq{}i,k,\
i\not=k\leq{}\left\vert{}S\right\vert{},1\leq{}j\leq{}\left\vert{}P\right\vert{},1\leq{}t\leq{}T-1
\label{eq:13}
\end{multline}
\vspace{-10mm}
\begin{multline}
x_{i,j,k,t+1}\leq{}\left\vert{}p_j\right\vert{}-\sum_{t^{'}=1}^tr_{i,j,k,t^{'}}\ \\ \forall{}1\leq{}i,k,\
i\not=k\leq{}\left\vert{}S\right\vert{},1\leq{}j\leq{}\left\vert{}P\right\vert{},1\leq{}t\leq{}T-1
\label{eq:14}
\end{multline}
During the migration process the servers must maintain the minimum availability threshold for each partition with constraint (\ref{eq:15}).
\begin{equation}
\sum_{i=1}^{\left\vert{}S\right\vert{}}z_{i,j,t}\geq{}A\ \  \forall{}1\leq{}j\leq{}\left\vert{}P\right\vert{},1\leq{}t\leq{}T
\label{eq:15}
\end{equation}
The total migration time is extended to include all migrations in progress in constraint (\ref{eq:16}) and stopping the migration process in constraint (\ref{eq:17}).
\vspace{-3mm}
\begin{multline}
w_t\geq{}x_{i,j,k,t}\ \\ \forall{}1\leq{}i,k,\
i\not=k\leq{}\left\vert{}S\right\vert{},1\leq{}j\leq{}\left\vert{}P\right\vert{},1\leq{}t\leq{}T
\label{eq:16}
\end{multline}
\begin{equation}
w_t\geq{}w_{t+1}\  \ \forall 1\leq{}t\leq{}T-1
\label{eq:17}
\end{equation}
\vspace{-3mm}
\subsection{Objective}
\vspace{-5mm}
\begin{equation}
minimize\left(\sum_{t=1}^Tw_t\right)
\label{eq:18}
\end{equation}
We minimize the total migration time in (\ref{eq:18}). As the optimization minimizes migration times, it will select source servers for replica migration that reduce migration time, such that it selects source-destination pairs that have minimum overlapping edges in the shortest path, while ensuring sequential migrations, meeting minimum partition availability threshold and abiding by edge bandwidth capacities.

\section{Solution Design}
\label{sec:SolutionDesign}

In this section we will describe CRANE, our heuristic solution for the replicas migration problem. Given an initial and a target replicas mapping in data centers, the goal of this algorithm is to find the best sources for copying the replicas and the best sequence to send them so as to minimize the total replica creation/migration time. To this end, the following sights can guide the replication replica creation/migration sequence : (1) avoid redundancy, (2) select the source of data and paths having more available bandwidth, and (3) avoid idle time between sequences.

Our  heuristic  solution  is  described  in  Algorithm 1.  Given an initial and target placement configurations (i.e., $C^I$ and $C^F$), Algorithm 1 returns a set $Q$ of sequences $Q_{i}$ for migration. Each sequence contains an ordered set of replicas to be migrated/created such that the required minimum availability per partition is satisfied. After each sequence of migrations $Q_{i}$, cloud storage components will be updated with the new placement, so that data user requests can be redirected to the right partition locations. The final replica placement is reached once all the sequences $Q_{i}, i<n$ are executed.

Initially, $P$ contains the set of partitions that need to be created/migrated. This set can be computed based on the initial and the final partition locations (i.e., $C^I$ and $C^F$). We then initialize $Q$ that should contain the sequence of replicas to be migrated. In line 3, we initialize a variable $i$ that denotes the number of the sequence. The core of the algorithm aims to iteratively add a partition replica on ordered sequence $Q_{i}$. We create a new sequence whenever it is not possible anymore to add a replica creation/migration to the current sequence (not possible because otherwise we do not satisfy the minimum replica availability per partition). 

The variable $available$ is $true$ if there still some replica that can be added to the sequence $Q_i$. 
As long as $available$ is $true$, we iterate over all partitions in $P$ and we choose the replica  $r_b$ from the set of replicas $R_p$ of each partition $p$ that minimizes the migration time. For that we use the variable $T_{Q_i,R_p,min}$ that denotes the minimal migration time of the sequence $Q_{i}$  when a replica of partition p is included. We compare this variable to $T_{Q_i,r}$, the migration time of the sequence $Q_i$ if replica $r$ is included (line 15). This allows us to select the best replica $r_b$ of the partition $p$. From the selected replicas, we need to select the one that minimizes migration time (denoted by $r_s$). To do so, we use the variable $T_{Q_i,min}$ which denotes the minimal migration time after adding a new replica to the sequence $Q_i$ (line 9). We then choose from all previously selected replicas the one that  minimizes migration time after adding a new replica (line 20 to 23). The chosen replica is then added to $Q_i$ (line 26). 
The partition $p$ whose replica $r_s$ was selected is then removed from the set $P$.

To detect that we cannot add any more replica to the sequence $Q_i$, we iterate over all partition until the variable $available$ becomes $True$. At that time, we add the sequence to $Q$, and start a new one as long as we still have partitions in $P$ to migrate/create.

\begin{algorithm}
\caption{CRANE}
\textbf{require:} Initial configuration $C^I$. \\
\textbf{require:} Final configuration  $C^F$ \\
 \textbf{output:} Sequence for migration \\
\begin{algorithmic}[1]
\STATE $P \leftarrow $ partitions to be migrated
\STATE $Q \leftarrow \{\emptyset\}$
\STATE $i \leftarrow 0$
\WHILE {$P \not=\{\emptyset\} $}
	\STATE $Q_{i} \leftarrow \{\emptyset\}$ 
	\STATE $available \leftarrow True$
	\WHILE {$available == True$}
		\STATE $available \leftarrow False$
		\STATE $T_{Q_i,min} \leftarrow	\infty	$
		\FOR{\textbf{each} $p \in P$}
			\IF{p can be included in $Q_{i}$}
				\STATE $available \leftarrow True$	
				\STATE{ $T_{Q_i,R_p,min} \leftarrow \infty$}
				\FOR{\textbf{each} $r \in R_{p}$}
 					 \IF{$ T_{Q_i, r} < T_{Q_i,R_p,min}$}
    						\STATE $T_{Q_i,R_p,min} = T_{Q_i, r}$
    						\STATE $r_{b}  = r$
   					\ENDIF
 				\ENDFOR
 				\IF{$T_{Q_i,R_p,min} < T_{Q_i,min}$}
 					\STATE $T_{Q_i,min} = T_{Q_i,R_p,min} $
 					\STATE $r_{s}  = r_{b}$
				\ENDIF
			\ENDIF
		\ENDFOR
		\STATE $Q_{i} =  Q_{i} \cup r_{s}$
		\STATE $P = P - \{$partition $p$ of replica $r_{s}\}$
	\ENDWHILE
	\STATE $Q = Q \cup Q_{i} $
	\STATE $i \leftarrow i+1$
\ENDWHILE
\STATE \textbf{return} Q
\label{Algo}
\end{algorithmic}
\end{algorithm}

\section{Performance Evaluation}
\label{sec:PerformanceEvaluation}

In this section, we compare the performance of Swift using CRANE for replica migration with the traditional Swift, with respect to migration time, amount of transferred data and partition availability. 

\subsection{Deployment scenarios}
Our evaluation environment consists of five data centers, each consisting of five storage servers. We use the NSFNet topology as a backbone network interconnecting the data centers \cite{NSFNET}. 
We use the standard Swift configuration stipulating that for each data partition, three replicas have to be created and placed according as-unique-as possible algorithm. Furthermore, swifts assumes that if 2 out of the 3 replicas are available, the data is assumed to be available (i.e., all user requests can be accommodated). Hence, the minimum required availability per partition is set to $2/3$.

In our simulations, we consider four scenarios as depicted in Table~\ref{scenarios}. Each having different number of partitions placed across the data center with different partition sizes. 
In the beginning of each experiment, we consider only 4 data centers. After that, a new data center is connected to the infrastructure, which triggers the placement algorithm in order to re-optimize the location of replicas. We then use whether swift combined with CRANE or the traditional swift to migrate or create new replicas.
For instance, in scenario 1, we originally have 512 partitions (i.e., 1536 replicas) distributed across the infrastructure. When the fifth data center is added, 656 replicas should be migrated or created across the fifth data centers (Table~\ref{scenarios}). 



\begin{figure*}[htb!]
\begin{center}
\subfigure[Migration Time (min)]{
\includegraphics[scale=0.33]{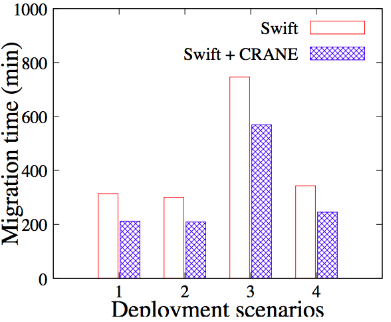}
\label{MigrationT}}
 \subfigure[Amount of transferred data (1000Gb)]{
 \includegraphics[scale=0.21]{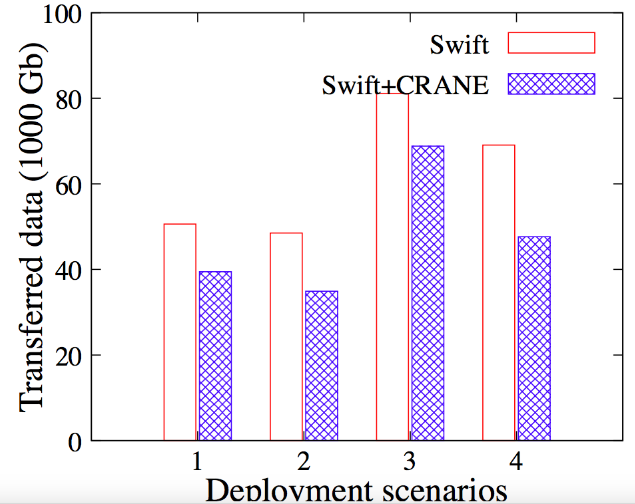}
\label{Size}}
 \subfigure[Availability]{
\includegraphics[scale=0.19]{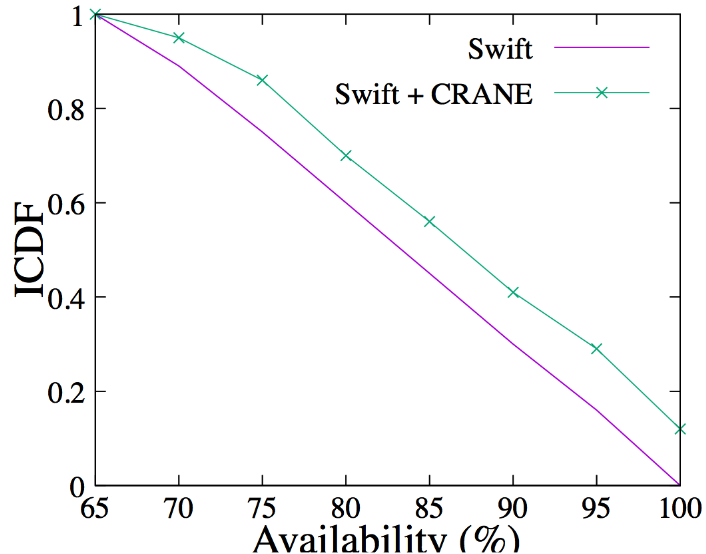}
\label{availability}
}
\centering
\caption{Performance comparison between CRANE and traditional Swift.}
\label{Results}
\end{center}
\end{figure*}


\begin{center}
\begin{table} [!tbh]
\caption{Deployment scenarios}
\begin{tabular}{|m{30pt}|p{55pt}|p{55pt}|p{55pt}|}
\hline
\textbf{Scenario} & \textbf{Total number of partitions} & \textbf{Number of replicas to migrate} & \textbf{Range of replicas size (Gb)}\\ 
\hline
1 & 512 & 656 & 50-100 \\
\hline
2 & 1024 &1316 & 20-50 \\
\hline
3 & 2048 & 2632 & 20-50 \\
\hline
4 & 4094 & 5264 & 10-20 \\
\hline
\end{tabular}
\label{scenarios}
\end{table}
\end{center}

\subsection{Results}
For each scenario, depicted in Table~\ref{scenarios}, we compare (Swift + CRANE) with traditional Swift with respect to the following performance metrics: (1) the total  migration  time, (2) the amount of inter-data centers exchanged data, and (3) the availability of replicas per partition.

Figure \ref{MigrationT} shows  the  total  migration time for the considered scenarios. As we can see, in all scenarios, (CRANE~+~Swift) outperforms  the  Swift  algorithm  by  a  good  margin.  For scenario 1, the swift algorithm takes 315 min to create all the replicas compared to 217 min for CRANE, which constitutes around 30\% of improvement. For scenario 2, the replicas are migrated within 300 min with swift and 200 min with CRANE, which constitutes around 30\% improvement. For scenarios 3 and 4, CRANE achieves 25\% improvement. These results are as expected, because CRANE always chooses to copy the replica incurring the minimal transmission time.

The amount of transferred data inter-data centers is  reported in figure \ref{Size}. For the 4 different scenarios the CRANE algorithm have less amount of data transferred. The improvement is around 25\%. This improvement is explained by the avoidance of redundant copy of the replicas of the same partition. This have also induced the improvement in migration time showed in figure \ref{MigrationT}. 

Finally, figure \ref{availability} shows the Inverse Cumulative Distribution Function (ICDF) of the availability. For a given availability, it provides the probability of having that availability or higher. 
The required minimum availability per partition ($2/3=0.66$)  is always met for both algorithms as we can see that the probability of having an availability higher than 66\% is 1. However, the probability of having a high availability is always higher for the CRANE algorithm than the traditional Swift. 
For instance, the probability of having an availability higher than 80\% is 0.60 for Swift whereas it is around 0.76 for CRANE. When comparing the two curves, we can see that, on average, CRANE improves availability by up to 10\%. 

It is clear that CRANE  performs  significantly  better  than  the basic Swift algorithm  as  it  carefully selects the replica from which the data should be copied, the paths used to transmit that data while avoiding redundant copy of replicas and eliminating idle time.

\section{Conclusion}
\label{sec:Conclusion}


Data replication has been widely adopted to improve data availability and to reduce access time. However, replica placement systems usually need to migrate and create a large number of replicas between and within data centers, incurring a large overhead in terms of network load and availability. 
In this paper, we proposed CRANE, an effiCient Replica migrAtion scheme for distributed cloud Storage systEms. CRANE complements replica placement algorithms by efficiently managing replica creation by minimizing the time needed to copy data to the new replica location while avoiding  network congestion and ensuring the required availability of the data. 
In order to evaluate the performance of CRANE, we compare it to the standard swift, the OpenStack project for managing data storage. Simulations show that CRANE is able to to reduce up to 30\% of the replica creation time and 25\% of inter-data center network traffic and provide better data availability during the process of replica migration.

\bibliographystyle{IEEEtran}
\bibliography{IEEEabrv,IEEEExampl}

\end{document}